\documentclass[10pt,prl,twocolumn,showpacs,aps,floats,floatfix,superscriptaddress]{revtex4}
\usepackage{latexsym}
\usepackage{graphicx}
\usepackage{amssymb}
\usepackage{epsfig}
\usepackage{amssymb}
\usepackage{amstext}

\begin{document}
\title{Microdynamics in stationary complex networks}

\author{Aurelien Gautreau}

\affiliation{Laboratoire de Physique Th\'eorique (CNRS UMR 8627),
Universit{\'e} de Paris-Sud, 91405 Orsay, France}

\author{Alain Barrat}

\affiliation{Laboratoire de Physique Th\'eorique (CNRS UMR 8627),
Universit{\'e} de Paris-Sud, 91405 Orsay, France}

\affiliation{Centre de Physique Th\'eorique (CNRS UMR 6207), Luminy Case 907,
13288 Marseille Cedex 9, France}

\affiliation{Complex
Networks Lagrange Laboratory, ISI Foundation, Torino, Italy}

\author{Marc Barth{\'e}lemy}

\affiliation{CEA-D\'epartement de Physique Th{\'e}orique et
  Appliqu{\'e}e, 91680 Bruyeres-Le-Chatel, France}

\affiliation{Centre d'Analyse et de Math\'ematique Sociales
  (CAMS, UMR 8557 CNRS-EHESS), Ecole des Hautes Etudes en Sciences
  Sociales, 54 bd. Raspail, F-75270 Paris Cedex 06, France}

\date{\today}
\begin{abstract}

  Many complex systems, including networks, are not static but can
  display strong fluctuations at various time scales. Characterizing
  the dynamics in complex networks is thus of the utmost importance in
  the understanding of these networks and of the dynamical processes
  taking place on them. In this article, we study the example of the
  US airport network in the time period $1990-2000$. We
  show that even if the statistical distributions of most indicators
  are stationary, an intense activity takes place at the local
  (`microscopic') level, with many disappearing/appearing connections
  (links) between airports. We find that connections have a very broad
  distribution of lifetimes, and we introduce a set of metrics to
  characterize the links' dynamics. We observe in particular that the
  links which disappear have essentially the same properties as the
  ones which appear, and that links which connect airports with very
  different traffic are very volatile. Motivated by this empirical study, we
  propose a model of dynamical networks, inspired from previous
  studies on firm growth, which reproduces most of the empirical
  observations both for the stationary statistical distributions and
  for the dynamical properties.

\end{abstract}

\maketitle



\noindent

\section{Introduction}

Despite the presence of stable statistical regularities at the global
level, many systems exhibit an intense activity at the level of
individual components, i.e.  at the `microscopic' level. An important
illustration of this fact was recently put forward by
Batty~\cite{Batty:2006} in the case of city populations. Indeed, even
if the population Zipf plots display negligible changes in time, the
same city can have very different ranks in the course of
history. Similarly, many other systems, in particular occurring in
human dynamics studies, present simultaneously stationary statistical
distributions and strong time fluctuations at the microscopic level,
with activity bursts separated by very heterogeneous time
intervals~\cite{Eckmann:2004,Barabasi:2005,Vazquez:2006,Gonzalez:2008}.
These systems thus challenge us with the fundamental puzzle which
consists in reconciling an important dynamical activity occurring at
the local level on many timescales and the emergence of stable
distributions at a macroscopic level which can be maintained even when
the external conditions are highly
non-stationary~\cite{Solomon:1999}. For instance, the dynamics of the
rank is not consistent with processes such as preferential attachment
\cite{bara02} where the rank is essentially constant in time.

These issues naturally apply to the case where complex systems are
structured under the form of large networks. In most recent studies,
these networks have been considered as static objects with a fixed
topology. However, their structure may in principle evolve, links may
appear and disappear. Such topological fluctuations have important
consequences: many dynamical processes take place on complex networks
\cite{mendes03,Newman:2003,Pastor:2004,BBVbook}, and a non trivial
interplay can occur between the evolutions of the topology and of
these dynamical processes. The structure of the network strongly
influences the characteristics of the dynamical processes
\cite{BBVbook}, and the topology of the network can simultaneously be
modified as a consequence of the process itself. In this framework,
recent studies have been devoted to simple models of coevolution and
adaptive networks
\cite{Zimmermann:2004,Gross:2006,Holme:2006,Nardini:2008,Gross:2008}.
Another illustration of the importance of taking into account the
dynamics of the network is given by concurrency effects in
epidemiology~\cite{Morris:1997}. Indeed, while a contact network is
usually measured at a certain instant or aggregated over a certain
period, the actual spread of epidemics depends on the instantaneous
contacts. In such contexts, it is thus crucial to gain insights into
the dynamics of the network, possibly by putting forward convenient
new measures and to propose possible models for it.

These considerations emphasize the need for empirical observations and
models for the dynamics of complex networks, which are up to now quite
scarce. In this paper, we study the case of the US airport network
(USAN) where nodes are airports and links represent direct connections
between them. It is indeed possible to gather data on the time
evolution of this network~\cite{BTS} (see also \cite{Correa} for a
study of the yearly evolution of the Brazilian airport network) which
represents an important indicator of human activity and
economy. Moreover, air transportation has a crucial impact on the
spread of infectious diseases \cite{Colizza:2006,Crepey:2008}, and it
would be interesting to include its dynamical variations in
large-scale epidemiological modeling. We first present empirical
measures on the dynamics of the USAN. In particular, we provide
evidence for the large-scale statistical regularity of many
indicators, and we also define convenient metrics that enable us to
characterize the small-scale dynamical activity. We then propose a
model, based on simple but realistic mechanisms, which reproduces most
empirical observations.

\medskip
\section{Empirical observations: stable statistical distributions in a
  fluctuating system}

We analyze data available from the Bureau of Transportation
Statistics~\cite{BTS}. These data give the number of passengers per
month on every direct connection between the US airports in the period
$1990-2007$. We limit ourselves to the period $1990-2000$ during which
the data collection technique is consistent. We obtain $12\times
11=132$ weighted, undirected \cite{note1} networks where the nodes are
airports, the links are direct connections, and the weights represent
the number of passengers on a given link during a given month. We
denote by $w$ the weight of a link, by $k$ the degree (number of
neighbors) of a node, and by $s$ its strength, equal to the sum of the
weights of the issuing links \cite{Barrat:2004} (in the
air-transportation case, the strength gives thus the total traffic
handled by each airport).

\begin{figure}[h!]
\vskip .5in
\begin{center}
\includegraphics[width=7.0cm]{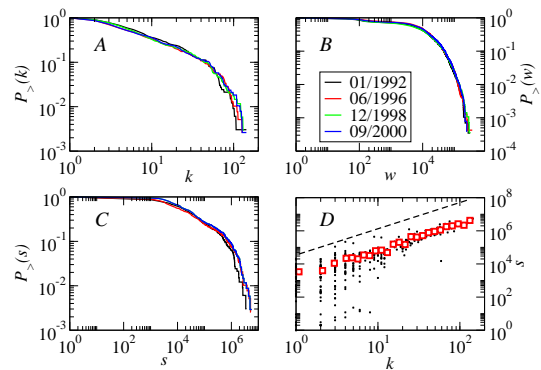}
\end{center}
\caption{ Characteristics of the US airport network
measured at four different dates: $01/1992$, $06/1996$, $12/1998$, and
$09/2000$. (A) Cumulative distribution of degrees; (B)
Cumulative distribution of weights; (C) Cumulative distribution
of strengths; (D) Strengths versus degrees for the year $2000$
(circles: raw data, squares: average strength for each degree value). The 
dashed line is a power law with exponent $1.6$.}
\label{fig1}
\end{figure}

Figure \ref{fig1} displays the cumulative distributions of degrees,
weights and strengths at four different times. These distributions are
broad, as already shown in previous studies
\cite{Barrat:2004,Guimera:2005,Barrat:2005}, highlighting the strong
heterogeneities present in the air transportation network, for both
the topology and the traffic. Figure~\ref{fig1}D moreover shows the
dependence of the strength $s$ of an airport on its number of
connections $k$, with a clear non-linear behavior denoting a strong
correlation between weights and topology
\cite{Barrat:2004,Barrat:2005}.  Interestingly, Figure \ref{fig1}
clearly shows that the distributions of degrees, weights, and
strengths measured at different times are identical (we have obtained
the same distributions at other dates). These distributions are
therefore stationary even if, as we will show later, non trivial
dynamics occur continuously.

\begin{figure}[thb]
\vskip .5in
\begin{center}
\includegraphics[width=7.0cm]{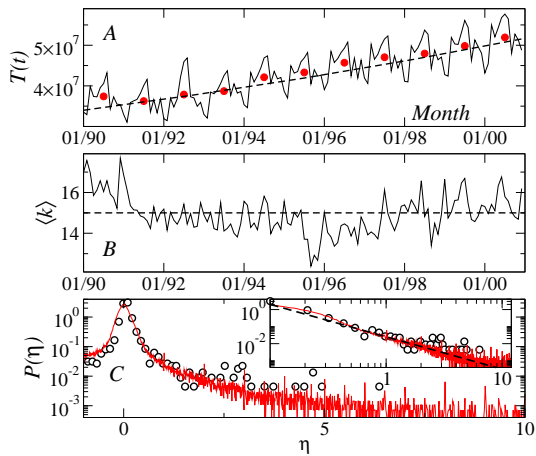}
\end{center}
\caption{(A) Total traffic on the US airport network versus
  time. The
  symbols represent the annual traffic and the dashed line is an
  exponential fit with time scale $\delta\approx 312$ months. (B)
  Average degree $\langle k\rangle$, 
approximately constant and of order $15$ (dashed line). (C)
  Distribution of the relative weights increments
  $\eta=(w(t+1)-w(t))/w(t)$. The full line corresponds to the
  distribution obtained over the $11$ years under study. Circles
  correspond to one single month (May $1995$). In the inset, we show
  the tail of the distribution of $\eta$, with a power law fit
  $P(\eta)\sim \eta^{-\nu}$, giving $\nu=1.9\pm 0.1$ (dashed line).}
\label{fig2}
\end{figure}
\begin{figure}[thb]
\vskip .5in
\begin{center}
\includegraphics[width=7.0cm]{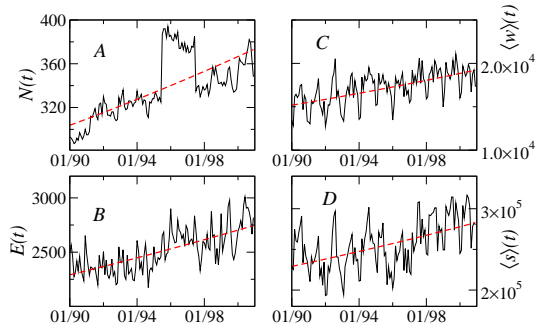}
\end{center}
\caption{Time evolution of (A) the number of nodes $N(t)$, (B) the
  number of links $L(t)$, (C) the average link weight $\langle
  w\rangle (t)$ and (D) the average node strength $\langle s\rangle
  (t)$ in the US airport network from January 1990 to December
  2000. Dashed lines are exponential fits.}
\label{fig:SI1}
\end{figure}

The first and simplest evidence for the presence of a dynamical
evolution in the network is displayed in Fig. \ref{fig2}A, which shows the
total traffic $T(t)$ (equal to the sum of the weights of all links) as
a function of time.
When the seasonal effects are averaged out, the data can be fitted, as
often assumed in economics, by an exponential growth
$T(t)=T(0)\exp(t/\delta)$ with $\delta\approx 312$ months
($\delta\approx 25$ years). Note that the data can also be fitted
linearly, due to the large value of $\delta$.  We also observe similar
growth (Figure~\ref{fig:SI1}) with seasonal fluctuations of the total
number $N(t)$ of connected airports, the total number $L(t)$ of links,
the average weight and the node strength. The fits give the same
growth rate for $N(t)$ and $L(t)$ (roughly half the growth rate of
$T(t)$), and the average degree $\langle k\rangle=2L(t)/N(t)$ has
small fluctuations ($\pm 3$) around a constant value ($\approx 15$),
as shown in Fig. \ref{fig2}B, over the $10$ years period under study, while the
average weight $\langle w\rangle$ grows exponentially with a typical
time of order $2\delta$.

\medskip\noindent
\section{The dynamics at the microscopic level}

We now study in detail  the dynamics at the microscopic
level, i.e. the evolution of single links. We denote by $w_{ij}(t)$
the weight of the link between nodes $i$ and $j$ at time $t$ (in
months) and by $\eta_{ij}(t)=[w_{ij}(t+1)-w_{ij}(t)]/w_{ij}(t)$ the
relative variation of $w_{ij}$ from one month to the next. Figure \ref{fig2}C
shows the distributions of $\eta_{ij}(t)$ for all links present in the
network both at $t$ and $t+1$, for all months in the $10$ years
dataset under study (period January $1990$-December $2000$), as well
as for a single month ($t=$ May $1995$). The fact that the
distributions can be superimposed leads to the conclusion that the
weights' evolution can be modeled by the form
\begin{equation}
w_{ij}(t+1)=w_{ij}(t)\left(1+\eta\right)
\end{equation}
where the multiplicative noise $\eta$ is a random variable
whose distribution does not depend on the link $(i,j)$ nor on the time
$t$. The inset of Fig.~\ref{fig2}C moreover shows that the distribution of
$\eta$ is broad, with a power law behavior $P(\eta)\propto\eta^{-\nu}$
(with $\nu\approx 1.9\pm 0.1$) for $\eta>0$. The broadness of this 
distribution indicates that most relative increments are small but that
sudden and large variations of the weights can be observed with a
small but non negligible probability. 

\begin{figure}[thb]
\vskip .5in
\begin{center}
\includegraphics[width=7.0cm]{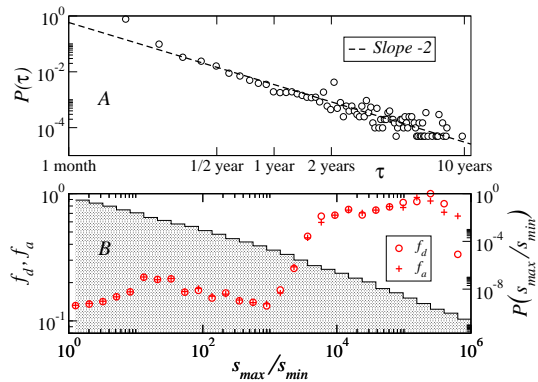}
\end{center}
\caption{(A) Lifetime distribution of links (in
months). Links which were existing at the beginning of the measure
(January $1990$) and still present at the end (December $2000$) are
discarded. The full line is a power law fit with an exponent
$-2.0\pm0.1$. (B) Fraction $f_d$ (open circles) of disappearing
links and $f_a$ (pluses) of appearing links as a function of the ratio
$s_{max}/s_{min}$ of the strengths of their extremities. The scale is on
the left-hand $y$-axis. We also show
the logarithmically binned
reference distribution $P(s_{max}/s_{min})$ (line above the shaded area, scale
on the right-hand $y$-axis).}
\label{fig3}
\end{figure}

The distribution of $\eta$ is truncated at $-1$, since this
corresponds to a weight going to $0$, i.e. to the disappearance of a
link. Links indeed can be created or suppressed between airports, and
in fact the number of link creation events is $4 \times 10^4$ for the
$11$ years period under study, for a total number of links in the
$132$ networks close to $3 \times 10^5$.  This result immediately
raises the question of the lifetime $\tau$ of links. As shown in
Fig.~\ref{fig3}A, the distribution of $\tau$ is very broad, with a power law
behavior $P(\tau)\sim \tau^{-\alpha}$ with $\alpha=2.0\pm0.1$. 
Some comments are in order. First, we consider in this distribution
only the links which appear {\em and} disappear during the period
under study. This is necessary since we cannot know the real lifetime
of a link which is already present at the start of the period or still
present at the end. Second, while the most probable value for $\tau$
is small, which implies that new links are the most fragile, the
distribution extends over all available timescales: links of an
arbitrary age may disappear. This indicates a non trivial dynamics
with appearance/disappearance of both `young' and `old' links. This
strong heterogeneity of lifetimes is in line with other results about
human activity~\cite{Barabasi:2005}, where it has been shown to have a
strong impact on dynamical processes \cite{Vazquez:2007}. It is
therefore important to characterize and incorporate it into models of
dynamically evolving complex networks.

\begin{figure}[thb]
\vskip .5in
\begin{center}
\includegraphics[angle=-90,width=7.0cm]{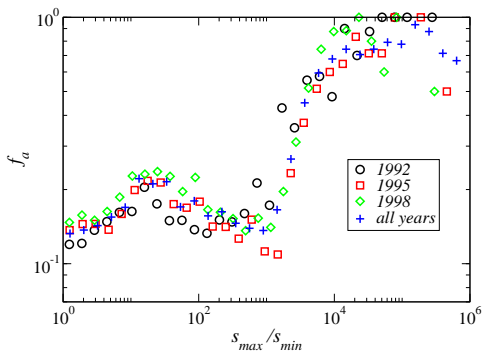}
\end{center}
\caption{Fraction $f_a$ of appearing links in the USAN as a
  function of the ratio $s_{max}/s_{min}$ of the strengths of their
  extremities. Circles, squares and diamonds correspond to the data
  of three distinct years, while the pluses represent the data
  averaged over the whole $10$-years time period. This figure clearly
  illustrates the stationarity of $f_a$.}
\label{fig:SI2}
\end{figure}

These results show that, behind the stability of the statistical
characteristics of the USAN, incessant microscopic rearrangements
occur. We now propose a systematic way to characterize the
corresponding fluctuating connections, whose importance stem from the
fact that they induce changes in the topology of the network.  Each
link $(i,j)$ can be characterized by a certain number of quantities
such as its weight $w_{ij}$, the strengths of its extremities $s_i$
and $s_j$, etc. It is usual to consider the distributions of these
quantities {\it over the whole network}, and we will consider these
distributions as reference (see Fig. \ref{fig1}).  In addition, we propose to
focus at each time $t$ on the links which appear (or disappear), to
study the distributions of these links' characteristics, and to
compare them with the reference distributions. For instance, if
$N_t(w)$ is the number of links with weight $w$ at time $t$, and
$N_t^d(w)$ is the number of such links that disappear between $t$ and
$t+1$, we measure the {\em fraction} of links of weight $w$ that
disappear at time $t$,
\begin{equation}
f_d(w)= \frac{N_t^d(w)}{N_t(w)} \ .
\label{eq:fd}
\end{equation}
We also define the number $N_t^a(w)$ and fraction $f_a(w)$ of links of
weight $w$ that appear at $t$. Similarly, $f_d$ and $f_a$ can be
measured for other links characteristics as we will investigate.  A
priori, all these quantities depend on the measurement time $t$. We
have already seen that the reference distributions are stationary
(Fig. \ref{fig1}). Strikingly, we observe that the fractions $f_d$ and $f_a$
display as well a stationary behaviour (Fig.~\ref{fig:SI2}), even if
they clearly highlight a strong dynamical evolution. In the following,
we will therefore drop any $t$ index and measure $f_d$ and $f_a$
averaged over the whole period under study.

The measure of $f_d(w)$ and $f_a(w)$ indicate that most links have a
small weight just before they disappear or just after their birth,
which is not a surprise. However, $f_d$ and $f_a$ are broad, extending
on several orders of magnitude of $w$ values: appearing and
disappearing connections occur with non negligible probabilities even
at strong weights. We also note a strong similarity between $f_d$ and
$f_a$, due to the large number of links with lifetime of order a few
months, during which no strong evolution of $w$ occurs.  A more
detailed analysis shows the presence of two regimes in $f_{d(a)}$: for
links with $w<10^2$ passengers per month, $f_d$ and $f_a$ present
rather large values close to $0.8$. For $w>10^2$, these fractions
decrease slowly: also links with large weights can appear or
disappear. We also observe that for $w>10^3$ there are more links
which appear than which disappear, an effect which is consistent with
the increase of the total traffic.

As previously mentioned, a similar analysis can be carried out for
various links' characteristics; particularly relevant quantities
include the traffic of the airports located at both ends of the link.
In the following we denote by
$s_{max}(l)=\max_{l=(i,j)}\left(s_i,s_j\right)$ and
$s_{min}(l)=\min_{l=(i,j)}\left(s_i,s_j\right)$ the larger and smaller
traffic of the extremities of a link $l=(i,j)$. A measure of the
importance of the link for $i$ and $j$ is given by $w/s_{min}$ and
$w/s_{max}$. For instance, if $w/s_{min}$ is small, the link carries
only a small fraction of $i$'s and $j$'s traffic; on the contrary, a
large $w/s_{max}$ indicates that the link is important for both its
extremities. The study of $f_{d(a)}(w/s_{min})$ and
$f_{d(a)}(w/s_{max})$ shows that most links which disappear/appear
display small values of these ratios, of order $w/s_{min}<10^{-3}$ and
$w/s_{max}<10^{-4}$. This means that most of these links have a small
importance for the airports to which they are attached. For larger
values of $w/s_{min(max)}$, the ratios $f_d$ and $f_a$ decrease, from
$\sim 0.7$ to $\sim 10^{-2}$, and surprisingly increase again (from
$\sim 10^{-2}$ to $\sim 10^{-1}$) for $w/s_{min}>10^{-1}$ and
$w/s_{max}>10^{-2}$. This phenomenon corresponds to links which are
very important for some airports, the extreme case being airports with
a single connection (these airports have thus usually a small
strength).

Finally, we also consider the ratio $s_{max}/s_{min}$ of the traffic
of the links extremities. This quantity indicates indeed how similar
the airports connected by the link are, in terms of traffic.  We plot
in Fig.~\ref{fig3}B the fractions $f_{d(a)}(s_{max}/s_{min})$ of links which
disappear (appear) as a function of $s_{max}/s_{min}$. On this figure
we also show the reference probability distribution
$P(s_{max}/s_{min})$ which displays a broad behavior: most links
connect airports of similar importance, but the ratio
$s_{max}/s_{min}$ varies over $6$ orders of magnitude, and a non
negligible fraction of links connect very different
airports. Interestingly, $f_{d(a)}(s_{max}/s_{min})$ 
displays two different regimes.
For $s_{max}/s_{min}<10^3$, small values of $f_d$ are obtained:
links which connect airports of similar, or not too dissimilar, sizes,
are rather stable. In the opposite case when $s_{max}/s_{min}>10^3$,
the fraction $f_d$ increases rapidly to reach another plateau, at
values of order $0.7-0.8$. This last regime corresponds to links
connecting airports with very different traffic, which turn out
to be the most fragile and to have a short lifetime.

We can now summarize the results of our empirical observations,
obtained through the analysis of the tools introduced in
Eq. \ref{eq:fd}: (i) The links which disappear have essentially the
same properties as the ones which appear. (ii) The
disappearing/appearing links have a weight which is low on average but broadly
distributed: large weights links may appear or disappear with a non
negligible probability. (iii) Most disappearing links have small
weights with respect to the traffic of their extremities, but links
appear or disappear in the whole range of $w/s$.  (iv) Links which
connect airports with very different traffic are very volatile. (v)
The lifetime of links is broadly distributed and covers all available
timescales.

The set of measures we have presented, while not exhaustive, is able
to give a clear characterization of the dynamics of the network under
study \cite{note2}. They are also easily applicable to any network
undergoing topological changes, and can be generalized to include
other links characteristics. 

The results of the empirical analysis may moreover serve as guidelines
in the elaboration of a model for dynamically fluctuating networks. In
particular and in contrast with most models found in the literature,
topological modifications of the network result here from the
stochastic evolution of weights.

\medskip\noindent
\section{A model for dynamical networks}

Using the results of the empirical analysis of the airport network as
guidelines, we now propose a model for dynamically fluctuating
networks able to reproduce the main features observed for the USAN,
and which highlights important features of dynamical networks
modeling. We consider simple ingredients that can easily be extended
with more detailed rules, and can therefore serve as a modeling basis
in many other fields where the dynamics of weights and links is
essential. In this model, topological modifications of the network
result from the stochastic evolution of weights.

We start from ideas developed in
\cite{Amaral:1996,Amaral:1998,Lee:1998} to model firm
growth through a process based on multiplicative growth of subunits
together with fusion/creation rules. In our framework, we consider
airports (nodes) and connections (links) instead of firms and
subunits. The equivalent of a firm's size is then given by the traffic
of the airport as measured by its strength, and the subunits sizes
correspond to the traffic on each link. An essential difference
distinguishes our model from the firm growth model where the various
firms undergo independent evolutions: here, each node is connected to
many others by links whose weights evolve randomly, so that the
evolution of the airports sizes are correlated.

Let us present the details of the modeling framework.  We start (at
time $t=0$) from an initial network composed of $N_0$ and $L_0$ links
with $L_0\approx N_0$ (we have checked that the initial conditions do
not influence the results). At each time step $t$, we first compute
for each link $(i,j)$ with weight $w_{ij}(t)$ a random increment
\begin{equation}
 \delta w_{ij}(t)=w_{ij}(t)\eta\ ,
\label{deltaw_model}
\end{equation} 
where $\eta$ is a random variable drawn from a distribution
independent from time and from the pair $(i,j)$, and which may a
priori take values in $]-1,+\infty[$. For $\langle \eta\rangle >0$,
    the total traffic will on average grow exponentially.  For the
    sake of simplicity we will choose for $\eta$ a Gaussian
    distribution (truncated at $-1$), with variance $\sigma^2$
    \cite{note3}. The weights' increments govern the evolution of the
    network's topology: depending on the values of $\delta w_{ij}$,
    the nodes $i$ and $j$ can either update the weight of $(i,j)$,
    delete it or create new links towards other nodes. More precisely,
    each airport $i$ has a threshold value $s_s(i)$ which sets a
    criterium of viability for a connection: if a link's weights drops
    below this threshold, the airport $i$ does not consider the link
    anymore as interesting and removes it. For simplicity, we take
    thresholds independent from time and uniform: $s_s(i)=1.0$ for all
    $i$.  The detailed evolution rules are as follows:
\begin{itemize}
\item (1) If $\delta w_{ij}(t)<0$, $i$ and $j$ test each the viability
  of the connection $(i,j)$. If $w_{ij}(t)+\delta
  w_{ij}(t)<\max\left(s_s(i),s_s(j)\right)$, the link disappears and
  its weight is uniformly redistributed over the other connections of $i$ and
  $j$. In the opposite case, $w_{ij}(t)+\delta
  w_{ij}(t)>\max\left(s_s(i),s_s(j)\right)$, the link's weight is
  simply updated: $w_{ij}(t+1)=w_{ij}(t)+\delta w_{ij}(t)$.

\item (2) If the weight increment $\delta w_{ij}(t)$ is positive, we
  assume that $i$ and $j$ have contributed equally to it and can
  decide each on how half of it should be used: If $\delta
  w_{ij}(t)>s_s(i)$, with probability $p_f$ node $i$ will use its part
  $\delta w_{ij}(t)/2$ of the increment to create a new link
  $(i,\ell)$ with weight $w_{i\ell}=\delta w_{ij}(t)/2$. With
  probability $1-p_d$, $\ell$ is an existing airport chosen at random,
  and with probability $p_d$ it is a new node.  $p_d$ therefore
  governs the rate of growth of the number of nodes. With probability
  $1-p_f$, node $i$ simply increases the weight $w_{ij}$ of $\delta
  w_{ij}(t)/2$.  Node $j$ then chooses independently either to create
  a new link $(j,k)$, or to increase the weight $w_{ij}$ by an amount
  equal to $\delta w_{ij}(t)/2$.

\item (3) If $0<\delta w_{ij}(t)<s_s(i)$, node $i$ increases the weight
  of $(i,j)$ of $\delta w_{ij}(t)/2$.
  The same procedure is applied to node $j$.

\end{itemize}

The rules (1)-(3) express the concept that the evolution of the
traffic governs the topological modifications of the network. If a
weight becomes too small, the corresponding connection will be
stopped. On the other hand if it grows too fast, new connections can
be created. The quantity $p_f$ determines the rate of new
connections. If $p_f$ is close to one, as soon as an increment $\delta
w$ is large enough a new link will be created, which in turn will
limit the growth of weights since they are used to create new
connections. In the opposite case of small $p_f$, the number of links
will grow very slowly but the weights will reach more easily large
values.  At each time step, the total traffic $T(t)$ is multiplied on
average by $1+\langle \eta\rangle$ leading to an exponential growth
$T(t)\propto \exp(t/d)$ with $d=1/\langle\ln(1+\eta)\rangle$. The
number of nodes and links also grow in time, and their simultaneous
growth, controlled by $p_f$ and $p_d$, results in an average degree
$\langle k\rangle$ which fluctuates around a constant value, function
of the parameters $p_f$, $\sigma$, $\langle \eta\rangle$, and
$p_d$. For instance, for larger $p_d$, $N(t)$ grows faster and
$\langle k\rangle$ is smaller.

The model rules can easily be modified to incorporate other elements,
such as preferential attachment mechanisms or random distributions of
the threshold values $s_s(i)$. While we will focus here on the
simplest version as described above, we have also considered variants
(i) in which the link's relevance is tested if $\delta
w_{ij}<\max\left(s_s(i),s_s(j)\right)$ (instead of the condition $\delta
w_{ij}<0$), or (ii) where the weight of deleted links is
re-distributed at random, or (iii) only one new link can be created,
either from $i$ or $j$. The  conclusion is that the
qualitative features are not modified, showing that the simulation results presented
below are robust with respect to such changes.
We have also simulated the case $p_d=0$ in
which no new nodes are inserted, $N(t)=N_0$. In this case, the global
increase of traffic leads at large time to a fully connected network,
but during a long time, it remains sparse ($\langle k \rangle \ll N(t)$)
and the same results are again obtained in this regime.

\begin{figure}[thb]
\vskip .5in
\begin{center}
\includegraphics[width=7.0cm]{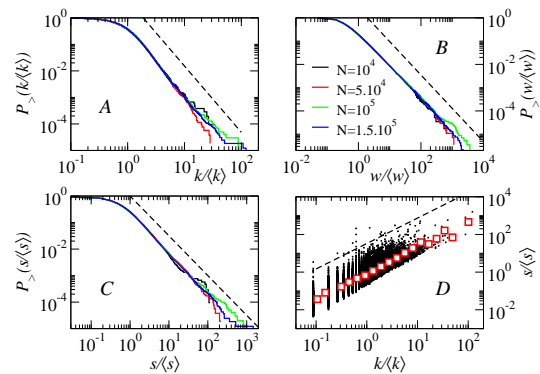}
\end{center}
\caption{Model simulation obtained with the parameters
  $p_d=0.01$, $p_f=0,1$, $\sigma=0.0225$ and $\langle
  \eta\rangle=10^{-3}$. Cumulative distributions obtained at different
  times of the network growth ($N=10^4$, $5.10^4$, $10^5$ and $1.5.10^5$)
  of: (A) normalized degrees $P(k/\langle k\rangle)$, the dashed
  line represents a power law with exponent $-2.5$; (B)
  normalized weights $P(w/\langle w\rangle)$, the dashed line
  represents a power law of exponent $-1.5$; (C) normalized
  strength $P(s/\langle s\rangle)$, the dashed line represents a power
  law with exponent $-1.5$. (D) Strength versus degree. Circles
  represent $s(k)$ for each node; full squares represent the same
  data, averaged for each $k$, and binned logarithmically.  This
  measure is done for $N=10^5$ and the dashed line represents a power
  law with exponent $1.4$.}
\label{fig4}
\end{figure}

\begin{figure}[thb]
\vskip .5in
\begin{center}
\includegraphics[width=7.0cm]{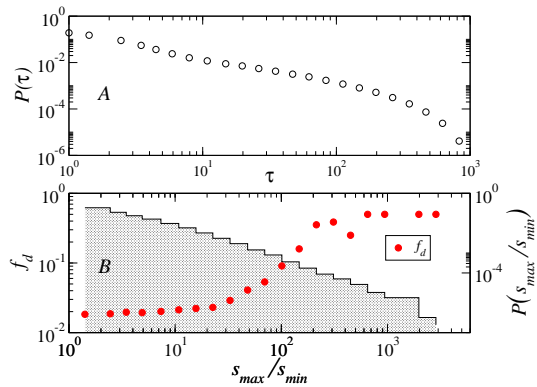} 
\end{center}
\caption{Model simulation with the same parameters
as in Fig. \ref{fig4}. (A) Lifetime distribution
$P(\tau)$ displaying a broad behavior. A power law fit gives an exponent of 
order $-1.1\pm 0.1$. (B) Fraction of
disappearing links $f_d$ versus $s_{max}/s_{min}$ (circles). We
also represent the logarithmically binned reference
distribution of $s_{max}/s_{min}$
(line above shaded ares).}
\label{fig5}
\end{figure}

Figures \ref{fig4} and \ref{fig5} summarize some results of our numerical simulations of
the dynamical network model. Although the network evolves with many
links creations and deletions, the distributions of degrees, weights,
and strengths display a remarkable stability, as shown in Figure \ref{fig4} for
$N(t)$ growing from $10^4$ to $10^5$. 
All these distributions are
broad, consistently with empirical observations, and the non-linear
behavior of the strength versus degree is reproduced as
well. Interestingly, this behavior (a power law with an exponent of
the order $1.4$, see Fig.~\ref{fig4}D) emerges here as a result of the
stochastic dynamics without any reference to preferential attachment
mechanisms combined with spatial constraints \cite{Barrat:2005} or
with link additions between nodes \cite{Bianconi:2005,Wang:2005}.

While many network models are able to produce broad degree and
strength distributions, the focus of this paper lies in the
small-scale dynamical aspects. We show in Fig. \ref{fig5}A that the lifetime
distribution of the links is broad, as in the USAN case, and we report
in Fig. \ref{fig5}B the behavior of $f_d(s_{max}/s_{min})$. Strikingly, our
model reproduces the empirical behavior shown in Fig. \ref{fig3}B, with two
different plateaus at small and large $s_{max}/s_{min}$. Other
properties of the appearing or disappearing links coincide in the
model with the empirical results, such as the fact that most
disappearing links have a small weight, or the non-trivial shape of
$f_{d(a)}(w/s_{min(max)})$, with a decreasing $f_d$ for $w/s>0.01$,
and an increase at $w/s>0.1$.

In summary, the simple assumptions on which our model is based yields
stationary non-trivial emergent properties such as broad distributions
and nonlinearities, together with an active local dynamics of links
occuring on all time scales, and whose characteristics reproduce the
empirical findings concerning the USAN's microscopic dynamics.

\medskip
\noindent
\section{Discussion}

The question of the dynamical evolution of networks is crucial in the
study of many dynamical processes and complex systems. If the time
scales governing the dynamics of the network and of the process taking
place on it are comparable, one can indeed expect a highly non trivial
behavior, which in principle could be very different from the static
network case. In this article, we have used as a case study the
US airline network, and we have shown that it exhibits stationary
distributions despite the incessant creation and deletion of
connections on broadly distributed timescales. We have introduced a
set of measures in a systematic way in order to characterize this
dynamics. Finally, we have proposed a model based on simple assumptions which
reproduces the main empirical features, both for stationary
and local dynamical properties. 

The coexistence of stationary distributions and strong microscopic
activity taking place at very different timescales occurs in many
different systems and our model can provide a framework that can
easily be extended and serve as a basis for further and more detailed
modeling. For instance, we have observed that a bimodal distribution
of the thresholds $s_s(i)$ for the deletion of a link results in the
following picture: nodes with small $s_s$ have typically a large
degree, but are connected to weak links, while nodes with large $s_s$
reach a smaller number of stronger connections. This behavior does not
correspond to infrastructure networks such as the USAN but could
describe social behavior where individuals with many connections do
not have intense (i.e. with large weight) relations. In these
perspectives, the present work should stimulate further studies on the
coexistence of dynamics at different scales and on the impact of
network dynamics on different processes.

\vspace{0.3cm}

\noindent{\bf Acknowledgements}: We thank V. Colizza and J.J. Ramasco for 
a careful reading of the manuscript and interesting suggestions.





\begin{thebibliography}{unsrt}
\bibitem{note1} From a rigorous point of view, the weights
are not fully symmetric, but the deviations are small so we will use the
symmetric undirected graph and avoid the complication deriving from
flow imbalances.
\bibitem{note2} Note that a straightforward other set of measures
  concerns the degrees of the links' extremities, see also
  \cite{Ramasco} for the definition of interesting measures combining
  weights and degrees. We have focused here on traffic properties,
  which are particularly relevant in the airport network case.
\bibitem{note3} We have also run simulations with broad distributions
  of $\eta$ for $\eta>0$, with the same results.
\end{thebibliography}
\end{document}